%% file: MWRingTrapPRLV4.0.tex
\documentclass[aps,pra,reprint,twocolumn, a4paper, superscriptaddress,showpacs]{revtex4-1}
\usepackage{bm}
\usepackage{enumerate} 
\usepackage[dvips]{graphics}
\usepackage[dvips]{graphicx}
\usepackage[dvips]{color}
\usepackage{pst-all}
\usepackage{epsfig}
\usepackage{amssymb}
\usepackage{amsmath}
\usepackage{flafter}
\usepackage{psfrag}                      

\usepackage{hyperref}

\begin{document}

\title{Inductively guided circuits for ultracold dressed atoms}

\author{German Sinuco-Le\'on}
\affiliation{Department of Physics and Astronomy, University of Sussex, Falmer, Brighton, BN1 9QH, United Kingdom}

\author{Kathryn Burrows}
\affiliation{Department of Physics and Astronomy, University of Sussex, Falmer, Brighton, BN1 9QH, United Kingdom}

\author{Aidan S. Arnold}
\affiliation{Department of Physics, SUPA, University of Strathclyde, Glasgow G4 0NG, United Kingdom
}

\author{Barry M. Garraway}
\affiliation{Department of Physics and Astronomy, University of Sussex, Falmer, Brighton, BN1 9QH, United Kingdom}

\date{\today}
\begin{abstract}
We propose a flexible and robust scheme to create closed quasi-one dimensional guides for ultra-cold atoms through the dressing of hyperfine sub-levels of the atomic ground state. The dressing field is spatially modulated by inductive effects over a micro-engineered conducting loop, freeing the trapping region from leading wires in its proximity. We show that arrays of connected ring traps can also be created by carefully designing the shape of the conducting loop. We report on characteristics of the trap and  mechanisms that limit the range of parameters available for experimental implementation, including non-adiabatic losses and heat dissipation by induced currents. We outline conditions to select appropriate parameters for operation of the trap with atom-chip technology. 

\end{abstract} 

\maketitle
Techniques for trapping and manipulating ultra-cold atomic matter on the micron scale have dramatically developed during the last two decades. In general terms, such fine control is possible thanks to precise temporal and spatial resolution of electric and magnetic fields, demonstrated in several experimental configurations that benefit from advances in micro-fabrication, electronic control, and laser technology \cite{RevModPhys.79.235}. These technical developments have lead to impressive experimental demonstrations of macroscopic quantum phenomena, such as matter wave interferometry \cite{schumm2005matter}  and persistent matter flux \cite{PhysRevLett.110.025302,PhysRevLett.110.025301}, and are at the heart of promising developments of technological applications in metrology \cite{0953-4075-43-11-115302}, quantum information technology \cite{PhysRevA.79.022316} and quantum simulators \cite{Advances_QuSim,NatureQuantumSimulator}.

Ring traps (and toroidal traps) are of particular interest because of the possibility they offer to study physical phenomena in a non-trivial geometry with true periodic boundary conditions, and to create atomic analogues of solid state electronic devices (e.g. \cite{0953-4075-43-11-115302}). Trapping of cold gases in such geometries has been demonstrated with a variety of experimental techniques, requiring control over optical fields \cite{PhysRevLett.110.025302, PhysRevLett.110.025301,PhysRevLett.110.200406,1367-2630-11-4-043030,PhysRevA.76.061404} or magnetic field distributions \cite{PhysRevLett.95.143201,PhysRevA.73.041606,PhysRevA.81.031402,1367-2630-14-10-103047}. In addition, there are several proposals for ring traps that rely solely on the field produced by current carrying conductors, being suitable to be implemented with atom-chip technology (e.g.~\cite{PhysRevA.81.031402,PhysRevA.80.063615}), in which feeding wires can break desirable symmetries. Such an effect can be mitigated by employing inductive coupling \cite{PhysRevA.77.051402},  which has been demonstrated in millimetre sized ring traps \cite{1367-2630-14-10-103047} and proposed for microscopic ring traps based on generalizing the radio-frequency dressing approach \cite{PhysRevLett.86.1195} to an inductive system \cite{arXiv:1310.2070}. 

In this contribution we show that highly configurable one-dimensional microscopic guides for ultra-cold atomic matter result from the response of an inductive loop to AC magnetic fields tuned near the atomic ground state \textit{hyperfine} splitting of alkali atoms. This trapping scheme is ideal for atomic coherent manipulation due to the negligible spontaneous emission associated with hyperfine levels of the atomic ground state \cite{PhysRevA.74.022312}. In addition, this proposal does not require sophisticated optical control and it is free from potential symmetry breaking current carrying wires in the vicinity of the trapping volume \cite{west:023115,PhysRevA.80.063615,PhysRevLett.87.270401,PhysRevA.74.033619,PhysRevA.75.063406}. In addition, the system can be designed to create multiply connected atomic circuits, e.g. arrays of connected ring traps, having in mind applications that benefit from matter-wave interferometry as in \cite{0953-4075-43-11-115302}.

\begin{figure}[!htb]
\centering
\includegraphics[width=8.54cm]{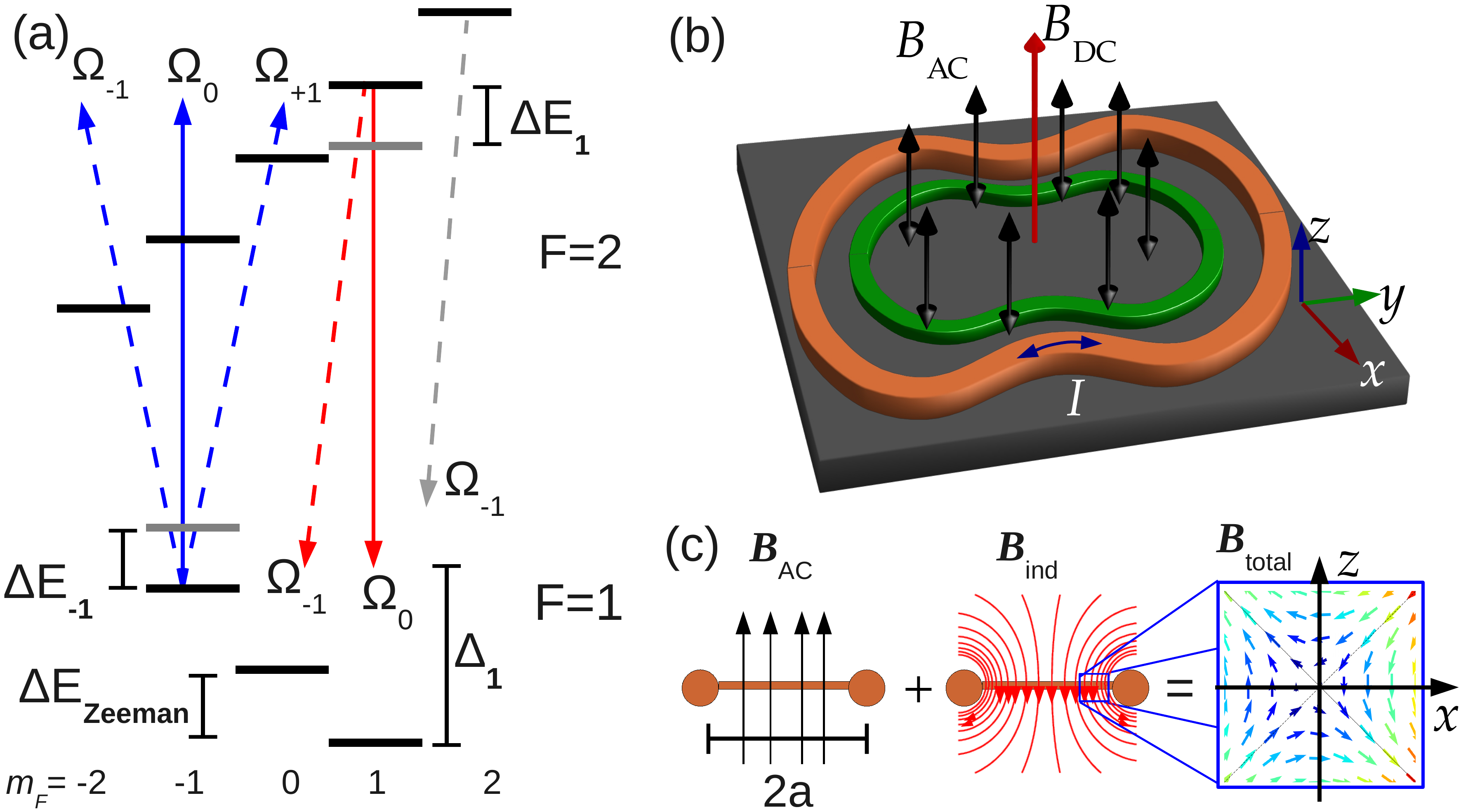}
\caption{\label{fig:figure1} (a) Ground state energy level structure of $^{87}$Rb. Arrows indicate magnetic dipole couplings between pairs of hyperfine sub-levels, corresponding to linear (solid lines)  and circular (dashed lines) polarizations of the magnetic field.  (b) Sketch of an atom-chip configuration to create an inductively coupled guide for ultra-cold atoms. It shows the magnetic field configuration (arrows), a closed conductor (orange) and the generated trapping region (green).  (c) Side view of the magnetic field distribution in the neighbourhood of the conductor: the uniform external field combines with the induced field and produces a total field with a quadrupole-like distribution.}
\end{figure}

For illustrative purposes,  we present calculations for the hyperfine level structure of $^{87}$Rb, denoted by $\left| F,m_F \right\rangle$, and shown in Fig. \ref{fig:figure1}(a). Nevertheless, our conclusions are straightforwardly extended to other atomic species with similar energy level structure. 

A sketch of the physical set-up is shown in Fig. \ref{fig:figure1}(b). It comprises a micro-engineered conducting loop (metallic or superconducting), a static magnetic field, $B_{\text{DC}} \hat{\boldsymbol{z}}$,  and a homogeneous AC magnetic field, $B_{\text{AC}}\cos(\omega t) \hat{\boldsymbol{z}}$, both transverse to the plane of the loop. In response to the electro-motive force induced by  the time variation of the magnetic flux across the area enclosed by the loop, an electric current circulates within it.  The induced current produces,  in its turn, an inhomogeneous  magnetic field of the form  $\boldsymbol{B}_{\text{ind}}(\boldsymbol{r}) \cos(\omega t + \delta)$, that modifies the total AC magnetic field.  For sufficiently large frequencies that the inductive reactance of the loop dominates its Ohmic resistance, the  external and induced fields are almost in anti-phase. Thus,  the resulting field has an approximately quadrupole distribution, schematically shown in Fig. \ref{fig:figure1}(c), whose centre is located close to the conducting loop at the position where the amplitude of induced and external fields satisfy $B_{\text{ind}} = B_{\text{AC}}\cos(\delta)$ \cite{PhysRevA.77.051402}, where $\delta + \pi$ is the relative phase between external and induced fields.

By tuning the driving frequency $\omega$ near the atomic ground state hyperfine transition, the AC magnetic field couples hyperfine Zeeman split sub-levels as depicted in Fig. \ref{fig:figure1}(a), leading to state-dependent potential energy landscapes for the atomic centre-of-mass motion  \footnote{Notice that we can safely ignore the electric field associated with the oscillating magnetic field, since the time-averaged quadratic Stark shift is proportional to the atomic DC polarizability of the ground state and thus independent of the quantum numbers $F$ and $m_F$ \cite{PhysRevA.74.022312}.}. The energy shifts are conveniently described in terms of the field components in spherical unit vectors $\hat{u}_{-1}=(\hat{x} - i\hat{y})/\sqrt{2}, \hat{u}_0=\hat{z}, \hat{u}_{+1} = -(\hat{x} + i\hat{y})/\sqrt{2}$, and corresponding Rabi frequencies $\Omega_{i} = \mu_B g_J B_i \left\langle F',m_F'\right| \hat{J}_i \left |F,m_F  \right\rangle$ with $i=-1,0,1$ and $g_J$ the Land\'e factor of the electronic angular momentum $J$. After the rotating-wave approximation and utilizing second order perturbation theory, near the quadrupole centre the energy shifts are given by \cite{PhysRevA.74.022312}
\begin{eqnarray}
 \Delta E_{m_F}(\boldsymbol{r}) &=& \pm \frac{1}{4}  \left(\frac{|\Omega_0 (\boldsymbol{r})|^2}{\Delta_{m_F}} \right.  \nonumber \\
&+ & \left. \frac{|\Omega_{-1} (\boldsymbol{r})|^2}{\Delta_{m_F} -\Delta_{\text{Zeeman}}}  +  \frac{|\Omega_{+1}(\boldsymbol{r})|^2}{\Delta_{m_F} + \Delta_{\text{Zeeman}}} \right)  
\label{eq:Energy_shift}
\end{eqnarray}
with $\Delta_{\text{Zeeman}} = \mu_B g_F B_{\text{DC}}$ and the detuning
\begin{equation}
\Delta_{m_F}  =  2A + \mu_B B_{\text{DC}} m_F (g_{F} - g_{F-1}) - \hbar \omega  \,.
\label{eq:detunning}
\end{equation}
where  the zero field hyperfine splitting of the ground state is $2A$, and $g_F$ the hyperfine Land\'e factor \footnote{We have checked that there are no significant differences between results obtained from Eq.\ (\ref{eq:Energy_shift}) versus a Floquet approach. This is because the trapping region is centred around a position of minimal field amplitude, where the perturbative expansion is valid.}. 

\begin{figure}[!htb]
\centering
\includegraphics[width=8.54cm]{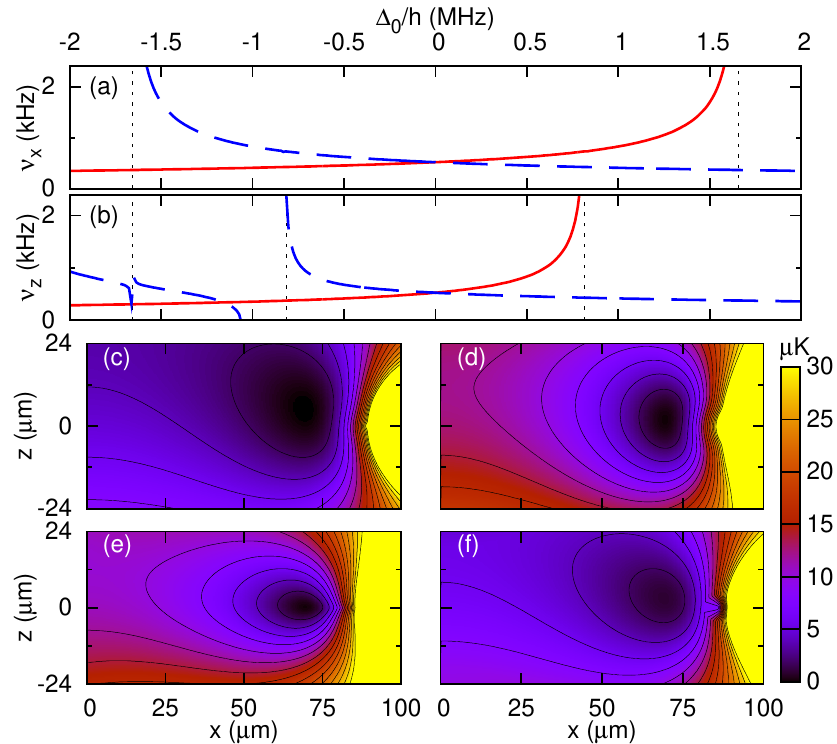}
\caption{\label{fig:figure2} (a,b) Trap frequencies corresponding to states $\left| F=2,m_F=1\right\rangle$ (solid) and $\left| F=1,m_F=-1 \right\rangle$ (dashed) of $^{87}$Rb, as function of the AC detuning,  with $B_{\text{AC}}=2$G, $B_{\text{DC}}= 1$G, along the (a) $x$ and (b) $z$ directions. Lower panels (c-f): Trapping potentials for $\Delta_{0} = -1.1$MHz (left column) and $\Delta_{0}=0.5$MHz (right column), for the states (c),(d) $\left|F=2,m_F=1\right\rangle$  and (e),(f) $\left|F=1,m_F=-1\right\rangle$. Gravitational attraction is included.}
\end{figure}

To give an explicit example of the potential landscape emerging from Eq. (\ref{eq:Energy_shift}) we consider a circular loop of gold with radius $a=100 \mu$m and diameter $s=10\mu$m, corresponding to approximate resistance $R\approx 0.26\Omega$ and inductance $L\approx 0.33$nH \cite{jackson1999classical}.  In this case, the total field distribution produces a circular trapping region with typical landscapes as shown in Figs.\ \ref{fig:figure2}(c)-(f), for states  $\left| F=2,m_F=1\right\rangle$  and $\left| F=1,m_F=-1 \right\rangle$ of $^{87}$Rb,  applied fields of $B_{\text{DC}}=1$G and $B_{\text{AC}}= 2$G.

The resulting quadrupole AC field distribution produces harmonic confinement, since the linear dependence of the field amplitude with the distance to the quadrupole centre translates into a quadratic variation of the energy shift in Eq. (\ref{eq:Energy_shift}). The tightness of the trap, quantified by the spatial curvature of the $\Delta E_{m_F}(\boldsymbol{r})$ along the $\hat{x}$ and $\hat{z}$ directions in units of frequency,  is shown in Fig. \ref{fig:figure2}(a)-(b) as function of the detuning of the driving field (see Eq. (\ref{eq:detunning})).  According to Eq. (\ref{eq:Energy_shift}), the trapping tightness increases arbitrarily by reducing the detuning with respect to pairs of transitions, resulting in the divergent  behaviour in Fig. \ref{fig:figure2}(a)-(b) (vertical dashed lines) at integer multiples of  $\Delta = |g_F\mu_B B_{\text{DC}}| \approx 0.7$MHz for  $B_{\text{DC}}=1$G. 

This trapping scheme provides confinement of atoms in two hyperfine states in overlapping regions. In our example of Fig. \ref{fig:figure2}, detuning in the range $ \Delta_{0} \in [-0.5,0.5]$MHz produce energy-shift landscapes for states $\left| F=2,m_F=1\right\rangle$ and $\left| F=1,m_F=-1\right\rangle$ with approximately equal curvatures for both states.  Even better, these two states experience exactly the same potential landscape for a driving field resonant to the hyperfine splitting, $\Delta_0= 0$. Note that the static magnetic field makes this resonant driving  to be blue (red) detuned with respect to coupling of states with $m_F=-1$ ($m_F=1$), as schematically shown by the solid arrows in Fig. \ref{fig:figure1}(a).

The detuning of the driving field also provides control over the shape of the trapping cross-section, as seen in the potential landscapes in Fig. \ref{fig:figure2}(c-f). This is because the relative weights of the terms in Eq. (\ref{eq:Energy_shift}) can be adjusted by changing the offset field and the driving frequency that determine $\Delta_{m_F}$.

\begin{figure}[!htb]
\centering
\includegraphics[width=8.54cm]{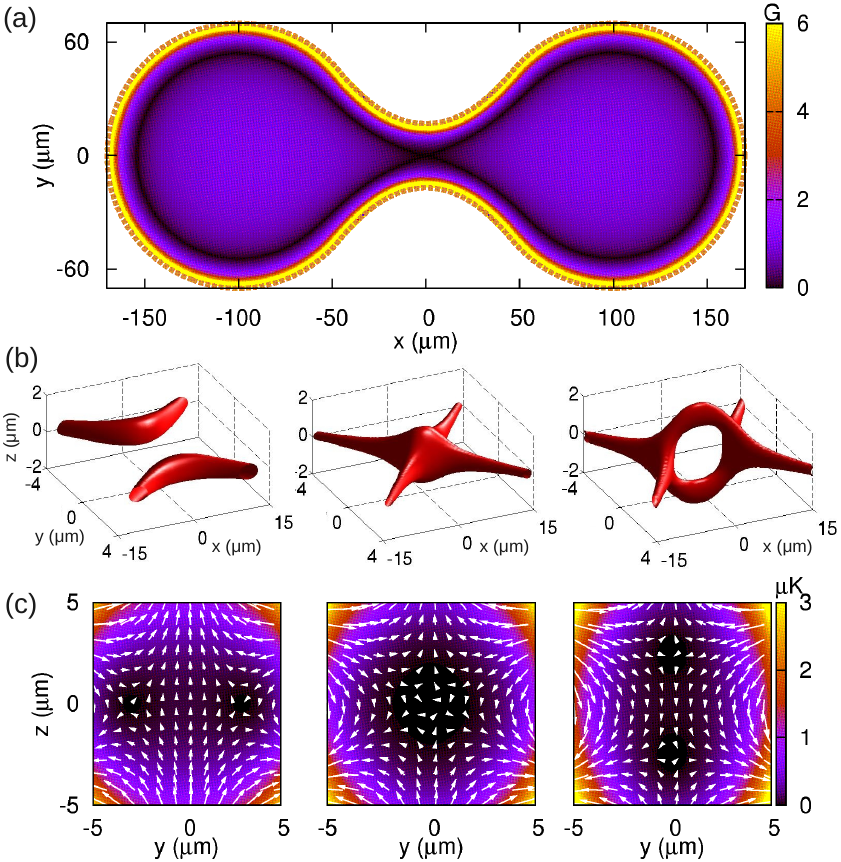}
\caption{\label{fig:figure3} A figure-of-eight  guide for atoms in the state  $\left|F=2,m_F=1 \right\rangle$ of $^{87}$Rb,  produced by a loop with a central symmetric constriction (orange dashed line in (a)). The conductor shape is defined by circles of radius $70\mu$m centred at $x=\pm 100\mu$m and a pair of parabolas that cuts the circle with matching first derivative.  (a) Magnetic field landscape in the loop plane, $z=0$, for the applied fields $B_{\text{DC}}=1$G, $B_{\text{AC}}=2$G. (b) Iso-energy surface at $0.5\mu$K  corresponding to central gaps of $35.2 \ \mu$m (left), $33.9\mu$m (centre) and $ 32.9\mu$m (right).  (c) Potential energy landscape and field distribution in the plane $x=0$, corresponding to surface plots directly above, in panel (b). In (b) and (c) $\Delta_{0}=0.35$MHz.}
\end{figure}

So far we have focused on the trapping produced by  a circular conductor. However, our scheme offers the possibility of creating complex atomic guides shaped by the conducting loop.  We illustrate this by considering a demanding case where we impose a severe `pinch' in the shape of the conducting loop, as depicted in Fig.~\ref{fig:figure3}, creating a double loop with a variety of junction geometries. The field distribution corresponding to this case can be understood as follows: away from the pinch centre, the field distribution is similar to the quadrupole field in Fig.~\ref{fig:figure1}(c), while in its neighbourhood the total field results from combining two quadrupole-like distributions associated with conducting segments at each side of the constriction.  In particular, when the induced field balances the applied one at the centre of the pinch, the field distribution acquires a hexapolar character. The geometry of the resulting potential landscape is sensitive to the shape of the conductor, while its energy scale is determined by the amplitude and detuning of the applied fields. This is illustrated in Fig.~\ref{fig:figure3}(b)-(c), where field distributions and energy landscapes have been obtained for three different constriction sizes differing by $\approx 1 \ \mu$m,  producing significantly different junction geometries. Consideration of this case can be straightforwardly applied to more complex geometries of the conductor, which can be used to create more involved atomic guides. 

\begin{figure}[!htb]
\centering
\includegraphics[width=8.54cm]{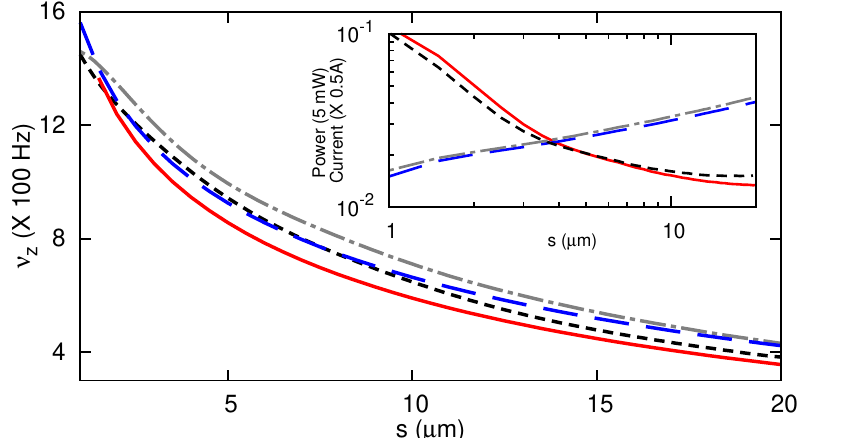}
\caption{\label{fig:figure4} Main panel: Trap frequency (in $100$Hz) as a function of conductor thickness ($s$), corresponding to circular and square cross-sections of superconducting Nb (solid and dashed lines), and gold (short-dashed and dot-dashed lines). Inset: Peak values of power dissipated (in factors of 5 mW, solid and short-dashed) and total current (in factors of $0.5$A, dashed and dot-dashed lines) in gold loops of circular (solid, dashed) and square cross sections (short-dashed and dot-dashed). Parameters as in Fig. \ref{fig:figure2} with $\Delta_{0}=0$.}

\end{figure}

Modelling the loop  as a  single current filament is insufficient to describe the potential landscape associated with conductors whose cross-section radius is comparable with the loop length \cite{1367-2630-14-10-103047,PhysRevLett.101.183006}. In such a case, the induced current distributes unevenly across the conductor and produces a magnetic field that differ significantly from the one produced by a single filament, having  direct impact on the quality of the trapping potential \footnote{See Supplemental Material for details on the method used to evaluate the current distribution within metallic and superconducting loops \cite{PhysRevLett.101.183006, tinkham1975introduction}.}. An illustration of these effects is shown in Fig.~\ref{fig:figure4}, where we consider circular loops with square and circular cross-sections made of two different conducting materials commonly used in atom-chip experiments: gold (Au) and superconducting niobium (Nb) \cite{PhysRevLett.101.183006}. 

In the case of a normal conductor,  the combination of small skin depth at high frequency with a radially dependent magnetic flux,  pushes the induced current towards the outer edge of the conductor, spreading the current along the conductor surface. Adding the Meissner effect according to the London description of superconductors \cite{tinkham1975introduction}, the current is confined even more dramatically in the case of superconducting loops.  As a consequence of distributing the current over a wide area, the gradient of the magnetic field is reduced in comparison to the single filament case. In terms of the atomic potential landscape, this translates to modifying the trapping position (i.e. the centre of the quadrupole field distribution) and reducing its tightness (here quantified through the trap frequency along the $x$ direction, $\nu_x$). Our numerical results indicate that both position and trap frequency, although dependent on  the conducting material and cross-section shape, do not vary strongly with these parameters. In both cases, the most relevant parameter is the thickness of the conductor, favouring the use of thin conductors to produce strong trapping potentials. 

The design of atom-chip configurations with current carrying elements is limited by several technical issues that restrict the range of experimentally accessible parameters \cite{RevModPhys.79.235}. In the present case, for example, the goal of obtaining the tightest possible trap, e.g. with small detuning or large driving fields,  should be balanced against an increase in heating and atom-loss rates.  In what follows, we briefly consider these two problems.

Ohmic loses due to the induced current must be restricted to avoid thermal destruction of the conductive loop,  or undesirable alteration of the trapping track due to thermal deformation of the conductor. For typical experimental parameters, such as those in Fig \ref{fig:figure2}, the average current densities (see inset Fig. \ref{fig:figure4}) are significantly lower than the maximal tolerable values demonstrated in experiments with normal and superconducting materials  operating under DC and high frequency conditions ($\approx 10^6$~A/cm$^2$) \cite{allcock2013microfabricated,PRA87_013436}, suggesting that the heat generated in our proposed trapping setup can be efficiently transferred to the supportive structures of the device. Also, although our numerical results for heating power favours using thick conductors, this should be balanced against the higher trapping frequency and better thermal coupling achievable with thin wires, which can support large current densities and are also convenient for fabrication \cite{RevModPhys.79.235}.  

We estimate non-adiabatic atom losses in our trapping setup by considering an atom moving at speed $u$ in the plane defined by the conducting loop. After the rotating-wave approximation, the atom-field interaction is described by the two-level  Hamiltonian \cite{PhysRevA.79.022316}: 
\begin{equation}
H =\frac{\Delta_{m_F}}{2} \sigma_z +  \frac{\Omega_{0}}{2} (\cos(\varphi)\sigma_x  + \sin(\varphi)\sigma_y)
\label{eq:Hamiltonian}
\end{equation}
where $\sigma_i$ with $i=x,y,z$ are Pauli matrices, and the spatially-dependent phase $\varphi$, and Rabi frequency $\Omega_{0}$,  are defined by the combination of the applied and induced fields. Atom-loss processes are modelled as transitions between the position-dependent eigenvectors of Hamiltonian Eq. (\ref{eq:Hamiltonian}), denoted by $\{\left|1\right\rangle,\left|2\right\rangle\}$ in the present treatment \cite{JOSAB_Tannoudji}.  Such dressed states consist of linear combinations for hyperfine states with the same projection of angular momentum $m_F$, that depend on the amplitude of the magnetic field.  For example, at the centre of the quadrupole field distribution, where the field is null,  the dressed states $\left| 1\right\rangle,\left| 2 \right\rangle$ coincide with the hyperfine states $\left| F,m_F\right\rangle,\left| F-1,m_F\right\rangle$, while very far from the zero they are an equal superposition of these two states. In the trapping geometry produced by a circular loop of inductance $L$ and radius $a$, the rate of transitions between pairs of dressed states is approximately \cite{JOSAB_Tannoudji}:
\begin{equation}
\Gamma_{\left|1\right\rangle \rightarrow \left|2\right\rangle} \approx \frac{1}{2\pi}\sqrt{\frac{2}{m}} \left(\frac{2L}{\mu_0 a^2}\right)^3 \left( \frac{\hbar u}{4} \right)^2 \frac{|\Omega_0|^3}{(\Delta_{m_F})^{9/2}}
\label{eq:NonAdb_transitions2}
\end{equation}
Under typical experimental conditions, e.g. an atom moving with speed  $u \approx 10$~mm/s (corresponding to a temperature of $1  \mu$K), and for the trap configuration presented in Fig. \ref{fig:figure2}, Eq. (\ref{eq:NonAdb_transitions2}) predicts  non-adiabatic transitions  with a rate of $\sim 10^{-5}$~s$^{-1}$, allowing enough time for manipulation of the trapped atoms.

Feeding the external field into the conducting loop presents another potential challenge.  However, in the case of $^{87}$Rb, and atoms with similar mass, the driving field should have a frequency in the GHz range, for which the near-surface field of a microwave co-planar cavity could be suitable \cite{PhysRevA.74.022312}. For the case of $^{6}$Li and light atoms, the driving frequency falls in the MHz range, where additional techniques can easily be employed \cite{RevModPhys.79.235}.

In summary, we have shown that complex one-dimensional guides for ultra-cold matter can be defined by inductive effects over metallic and superconducting loops. For operation, the loop should be fed with a magnetic field that oscillates near resonance to the hyperfine splitting of the atomic ground state, which induces an electric current on the conductor without the need of leading wires that might introduce undesired features in the potential landscape. Our numerical  investigations indicate that experimental realization of this type of trap is realistic with current technology, predicting trapping frequencies varying from a few hundred  Hz to a few kHz. Interestingly, our scheme can produce overlapping trapping regions for two different hyperfine states, which might be of interest for atomic species where a low magnetic field Feshbach resonance is available, such as in $^{6}$Li, as well as complex quasi-one dimensional circuits for cold matter. 

We acknowledge fruitful comments and input from Brage Gording, David Lucas, Michael K\"ohl and Peter Kr\"uger.  This work was supported by EPSRC grant EP/I010394/1.
\bibliographystyle{apsrev4-1}
\bibliography{MWRingTrapV3,revtex-custm}

\include{MWRingTrapPRL_SM}

\end{document}

%% file: MWRingTrapPRL_SM.tex
\section*{SUPPLEMENTARY INFORMATION}

\section*{Induced currents in conductors with finite cross-section}

In this section we provide more information about the induced current within conducting rings in the setup schematically shown in Fig. \ref{fig:figure1}(b) of the main text. We focus on oscillating magnetic fields with an associated wavelength ($\lambda = c/\omega$) much larger than the dimension of the ring, and apply a quasi-static approximation to the Maxwell equations for the electromagnetic field \cite{tinkham1975introduction, jackson1999classical}. In Sec. \ref{SS:Copper} we present results for the current distribution in metallic rings taking  parameters corresponding to gold. In Sec. \ref{SS:Nb} we detail a procedure to evaluate the current distribution in rings described by the  London theory of superconductivity, with parameters corresponding to superconducting Niobium, adapting results from references \cite{PhysRevLett.101.183006} and \cite{PhysRevB.69.184509}.

\begin{figure}[!htb]
\includegraphics[width=8.54cm]{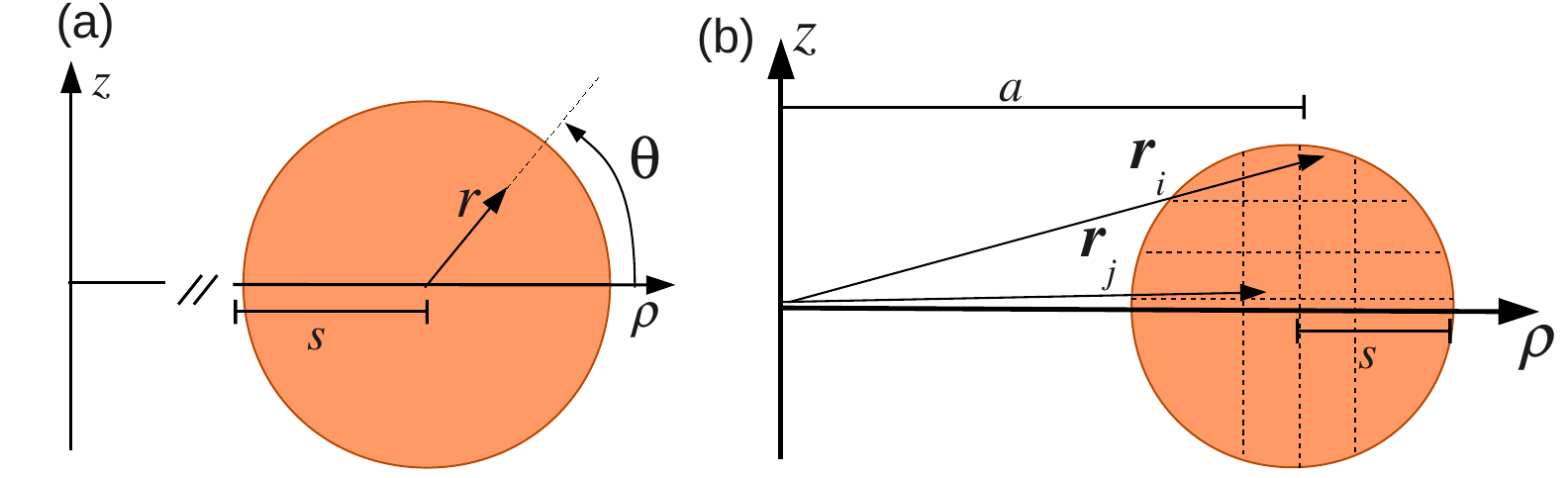}
\caption{\label{fig:figure5} (a) The current density in circular rings are evaluated at points defined by the polar coordinate system with origin at the centre of the conductor cross-section. (b) Cross-section of the coordinate system defined to evaluate the current distribution in conducting rings. For this work, we consider conductors with rotational symmetry around the $z$ axis. In both panels, the circular region represents the conductor cross-section.}
\end{figure}

We evaluate the current distribution using the coordinate systems in Figs. \ref{fig:figure5}(a)-(b). Exploiting the circular symmetry of the ring cross-section, the current density is evaluated at points defined by the polar-coordinate system with origin at its centre, as shown in Fig. \ref{fig:figure5}(a).   We express the  Maxwell equations coupled to a constitutive relation between the fields and the current in the ring (Ohm and London equations for metallic and superconducting materials, respectively) in the cylindrical coordinate system with origin at the centre of the ring, as defined in Fig. \ref{fig:figure5}(b).
 
\subsection{Metallic rings}
\label{SS:Copper}

The time-variation of magnetic flux across a metallic conductor induces an electric current whose distribution depends on the properties and  geometry of the ring as well as the frequency of the field.  For a harmonic variation of the magnetic field with frequency $\omega$, the quasi-static Maxwell equation for the vector potential is:

\begin{equation}
\nabla \times \nabla \times \boldsymbol{A} = - i\sigma \omega \boldsymbol{A} 
\label{eq:Maxwell-1}
\end{equation}
where $\sigma$ is the ring conductivity \cite{jackson1999classical}. 

\begin{figure}[!htb]
\includegraphics[width=8.54cm]{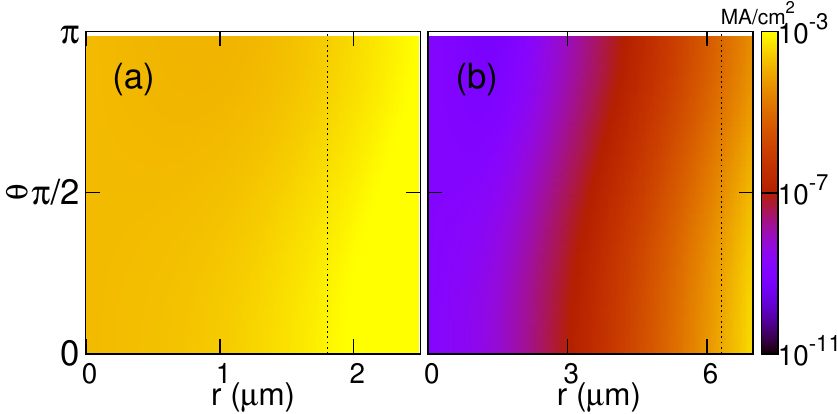}
\caption{\label{fig:figure6} Current density distribution across the cross-section of conducting loops of gold with thickness  (a) $2.5\mu$m and (b) $7\mu$m. In both cases, the ring radius is $a=100 \mu$m, $B_{\text{AC}} = 2$G and $\omega = 2 \pi \times 6.7$GHz. The vertical dashed line indicates the position of the skin-depth. 
}
\end{figure}

We use the open-source software package FEMM \cite{femm} to solve Eq. (\ref{eq:Maxwell-1}) for rings of gold with a range of cross-section sizes, under the action of a magnetic field oscillating at a frequency $\omega = 6.7$GHz. Figure \ref{fig:figure6} shows the current distribution for $R=2.5\mu$m and $R=7\mu$m. In rings of size comparable to the skin-depth at high frequencies, the current distributes across the hole area of the cross-section. In the case of large rings, the current concentrates along the conductor surface leaving the conductor centre free from current flow.   This confinement of the current impacts the power dissipated by the electric flow, as shown in Fig. \ref{fig:figure4} of the main text. 

FEMM also provides us results of the magnetic field distribution (not shown) which is then used to evaluate the trapping frequencies displayed in Fig. \ref{fig:figure4} of the main text.

\subsection{Superconducting rings.}
\label{SS:Nb}
We consider superconducting rings of uniform cross-section, described by the London theory \cite{tinkham1975introduction}, where the supercurrent and the potential vector are related by: 
\begin{equation}
\boldsymbol{J}(\boldsymbol{r}) =  -\frac{e^2 n_s}{m} \boldsymbol{A}(\boldsymbol{r})
\label{eq:LondonEq2}
\end{equation}
where $m$ and $e$ are the electron mass and charge, respectively, and $n_s$ is the density of superconducting electrons. Using this expression implies neglecting non-local effects on the current distribution as well as restricting the frequency of the oscillating field to values smaller than the superconducting gap (typically of the order of a few $\sim 100$GHz) \cite{tinkham1975introduction}.

In the presence of an external field and a given current distribution, under quasi-static conditions, the total vector potential is:
\begin{equation}
\boldsymbol{A}(\boldsymbol{r}) = \boldsymbol{A}_{AC}(\boldsymbol{r})  + \frac{\mu_0}{4 \pi}\int_{V}dV' \frac{\boldsymbol{J}(\boldsymbol{r'})}{|\boldsymbol{r}-\boldsymbol{r'}|}
\label{totalAfield}
\end{equation}
where the integral is over the volume of the current-carrying conductors.  $\boldsymbol{A}_{\text{AC}}$ is the vector potential associate with the applied field, which, in the case of a uniform magnetic field along the $z$ axis is $\boldsymbol{A}_{\text{AC}} =\hat{\boldsymbol{\phi}} \rho B_{\text{AC}}/2 $, imposing the Coulomb gauge condition $\nabla \cdot \boldsymbol{A}_{\text{AC}} = 0$ \cite{tinkham1975introduction}.

Superconducting rings with homogeneous cross-section have a current distribution independent of the azimuthal angle $\phi$,  and flowing tangentially to the perimeter of the conductor, i.e., along the direction $\hat{\boldsymbol{\phi}}$. This symmetry argument and Eq. (\ref{eq:LondonEq2}) allow us to simplify Eq. (7) to:
\begin{equation}
\frac{\rho B_{AC}}{2} \hat{\boldsymbol{\phi}}= \int dV' \hat{\boldsymbol{\phi}'} J(\rho',z') \left\{ \frac{m}{e^2 n_s} \delta(\boldsymbol{r}-\boldsymbol{r}')  +   \frac{\mu_0}{4 \pi}\frac{1}{|\boldsymbol{r} - \boldsymbol{r}'|} \right\}  
\label{eq:totalA2}
\end{equation}
where we have used an elementary property of the Dirac delta distribution \cite{PhysRevB.69.184509}. 

It is convenient to separate the integral over the volume of the conductor into an integral over the conductor cross-section and one over its circumference (see Fig. \ref{fig:figure5}(b)):
\begin{equation}
  \int dV'\hat{\boldsymbol{\phi}'} = \iint d\rho' dz' \times \oint_{\text{Ring}} d \boldsymbol{\ell'}
\end{equation}
where $d\boldsymbol{\ell'} = \rho' d\phi' \hat{\boldsymbol{\phi}'}$. Thus Eq. (8) becomes:
\begin{equation}
  \frac{\rho B_{AC}}{2} \hat{\boldsymbol{\phi}} =\int d\rho' dz'  J(\rho',z') \oint_{\text{Ring}} Q(\boldsymbol{r},\boldsymbol{r}') d\boldsymbol{\ell'} 
\label{eq:totalA2}
\end{equation}
with $Q(\boldsymbol{r},\boldsymbol{r}')$ defined as:
\begin{equation}
Q(\boldsymbol{r},\boldsymbol{r}') = \frac{m}{e^2 n_s} \delta(\boldsymbol{r}-\boldsymbol{r}')  + \frac{\mu_0}{4\pi}  \frac{1}{|\boldsymbol{r} - \boldsymbol{r}'|}
\end{equation}

Equation (8) can be recast in terms of magnetic flux across the loop $C$ defined by $\{\boldsymbol{r} = (\rho,\phi,z) \mid \phi \in [0,2\pi)\}$, using the relation $\Phi_C = \oint_C \boldsymbol{A} \cdot d \boldsymbol{\ell}$: 

\begin{equation}
\pi \rho^2 B_{\text{AC}} = \int d\rho' dz'  J(\rho',z') \oint_C \oint_{\text{Ring}} Q(\boldsymbol{r},\boldsymbol{r}') d\boldsymbol{\ell'}\cdot d\boldsymbol{\ell} 
\label{eq:totalA3}
\end{equation}

This equation implies that the magnetic flux across the loop $C$, created by the current distribution, compensates exactly the magnetic flux imposed by the external field. This corresponds to the well known Meissner effect in superconductors, and implies that the induced current adjust instantaneously in order to null the total flux of magnetic field across any loop defined within the superconducting ring. 

To obtain a solution of Eq. (12), we discretize the conductor cross-section in elements of area $\Delta A_i $ centred at positions $\boldsymbol{r}_i$, as schematically shown in Fig. \ref{fig:figure5}(b). Then, we obtain the equation:
\begin{equation}
\pi \rho_i^2 B_{\text{AC}}  = \sum_j L_{i,j} I_j 
\label{eq:totalA4}
\end{equation}
where $I_j=J_j \Delta A$, is the current flowing in the $j$-th loop, and:
\begin{equation}
L_{i,j} = \oint \oint Q(\boldsymbol{r}_i,\boldsymbol{r}_j) d\boldsymbol{\ell_i}\cdot d\boldsymbol{\ell_j}
\label{eq:InductanceMatrix}
\end{equation}
is the mutual inductance between the $i$-th and $j$-th loops, which for $i\ne j$ becomes:

\begin{equation}
L_{i,j} = \frac{\mu_0 \rho_i \rho_j}{4 \pi} \int_{0}^{2\pi} du \frac{\cos u}{( \rho_i^2 + \rho_j^2 + (z_i-z_j)^2 - 2 \rho_i \rho_j \cos u)^{1/2}}
\end{equation} 
This last integral is evaluated following \cite{PhysRevB.69.184509}.

For the self-inductance $L_{i,i}$ we follow \cite{PhysRevLett.101.183006}:
\begin{equation}
L_{i,i}=\mu_0 \rho_i \left[ \log \left(\frac{8 \rho_i}{R} \right)  - \frac{7}{4}\right] + \mu_0 \lambda^2 \frac{2 \pi \rho_i}{\Delta A_i}
\end{equation}
which includes the kinetic inductance term with $\lambda^2  = \frac{m}{\mu_0 n_s e^2}$.

In this work we consider superconducting ring of size $a=100 \mu$m, and circular cross-section in the range $s\in [1,20] \mu$m.  For Niobium, the London penetration depth $\lambda \approx 100$nm \cite{PhysRevLett.101.183006}.

Figure 7 presents the current distribution in rings with $s=2.5\mu$m and $s=7\mu$m, for an applied field $B_{\text{AC}}=1$G. In comparison to the case of metallic conductors shown in Fig. \ref{fig:figure6}, the current distribution concentrates more strongly near the surface of the conductor. Nevertheless, the impact on the trapping properties of the setup in Fig. \ref{fig:figure1} of the main text is similar in both cases, as shown in Fig. \ref{fig:figure4} also of the main text.

\begin{figure}[!htb]
\includegraphics[width=8.54cm]{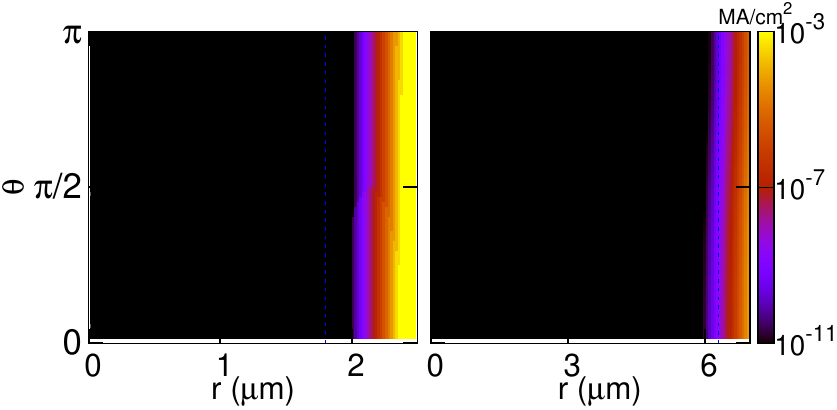}
\caption{\label{fig:figure7} Current density distribution across the cross-section of superconducting loops of Niobium with thickness  (a) $2.5\mu$m and (b) $7\mu$m. In both cases, the ring radius is $a=100 \mu$m, $B_{\text{AC}} = 2$G. The vertical dashed line indicates the position of the skin-depth. 
}
\end{figure}